\documentclass[aps,prl,twocolumn,superscriptaddress,floatfix,reprint]{revtex4-1}

\usepackage{epsfig,amsmath,amssymb,latexsym}

\usepackage{graphicx}
\usepackage{subfigure}
\usepackage{soul}
\usepackage{color,xcolor,xspace}
\usepackage[colorlinks,
            linkcolor=navy,
            anchorcolor=navy,
            citecolor=navy,
            urlcolor=navy
            ]{hyperref}
\definecolor{red}{rgb}{1,0,0}
\definecolor{blue}{rgb}{0,0,1}
\definecolor{navy}{RGB}{0,0,128}
\definecolor{black}{rgb}{0,0,0}

\begin{document}

\title{Identification of a Low Energy Metastable 1$T$-Type Phase for Monolayer VSe$_2$} \preprint{1}

\author{Bing-Hua Lei}
 \affiliation{Department of Physics and Astronomy, University of Missouri, Columbia, Missouri 65211-7010, USA}

\author{David J. Singh}
 \email{singhdj@missouri.edu}
 \affiliation{Department of Physics and Astronomy, University of Missouri, Columbia, Missouri 65211-7010, USA}
 \affiliation{Department of Chemistry, University of Missouri, Columbia, MO 65211, USA}

\date{\today}

\begin{abstract}
Elucidating the polymorphism of transition metal dichalcogenide layers and the interplay between structure
and properties is a key challenge for the application of these materials.
We identify a novel low energy metastable phase of monolayer VSe$_2$ and elucidate its magnetic
and electronic properties.
This structure is distinct from the previously identified charge density wave (CDW) phase.
However, while having rather distinct properties it is very close in energy to the CDW phase and is likely to
be realized in experiments.
Importantly, local bonding instabilities, as characterized by reconstruction of the electronic structure
over a wide energy range, are important for this distortion, which includes both V off-centering
in the octahedral coordination cages and a partial disproportionation into two distinct types of V.
The phase does not have a ferromagnetic ground state.
The results show that the physics of 1$T$ VSe$_2$ are richer than previously known with an interplay of
Fermi surface instabilities and local bonding effects.
\end{abstract}

\maketitle

Thin films and layers of transition metal dichalcogenides (TMDs) are an experimentally
accessible platform for realizing a wide variation of physical properties often very different from bulk
\cite{Manzeli_2017,Chhowalla_2013,van_Wezel_2011,Chatterjee_2015,Flicker_2015,zhang-CrTe2,ohara}.
These include band structure changes \cite{Mak_2010}, charge density waves (CDWs)
\cite{Strocov_2012,Pandey_2020,Chen_2018,Feng_2018,Si_2020,Xi_2015},
magnetism \cite{Bonilla_2018,Yu_2019,zhang-CrTe2,meng-CrTe2},
superconductivity \cite{Tsen_2015}, and various topological and quantum electronic states \cite{Yu_2015,Li_2015}.
These materials typically show a strong interplay between structure and properties. This includes
the very different properties of the 1$T$ (octahedral)
and 2$H$ phases (trigonal prismatic), and also an often strong interplay between properties
and different polymorphs from distortions of the 1$T$ and 2$H$ structures. The strong interplay between
properties and structure in these materials motivates identification of polymorphs and elucidation of their
properties \cite{duerloo,huang-rev}.

VSe$_2$ is a TMD of particular interest due to the observation high temperature ferromagnetism
\cite{Bonilla_2018,Yu_2019},
combined with charge density waves in the 1$T$ structure.
This ferromagnetism is in contrast to bulk 1$T$-VSe$_2$, which is a paramagnetic metal
\cite{Bayard_1976,van_Bruggen_1976}. 
The origin of this ferromagnetism and especially its intrinsic or extrinsic nature
are not yet established
\cite{Chen_2018,Feng_2018,Wong_2019,Yu_2019,Zhang_2019,Boukhvalov_2020,Chua_2020,Vinai_2020}.
Generally, both the nesting that leads to CDW instabilities and the density of states, which
may relate to magnetism \cite{he}, can be expected to be enhanced with dimensionality reduction, specifically
going from bulk to monolayers.

Bulk 1$T$ VSe$_2$ has a CDW instability
\cite{Bayard_1976,van_Bruggen_1976,Eaglesham},
indicating strong electron-phonon interactions.
While it is not superconducting at ambient pressure,
it does superconduct at high pressure where the CDW is suppressed
\cite{sahoo}.
The related 4$d$ and 5$d$ TMDs, NbSe$_2$ and TaSe$_2$ also show CDW transitions and superconductivity,
with an interplay between structure, CDW transitions and superconductivity \cite{luo}.
NbSe$_2$ is additionally superconducting in monolayer form \cite{wang}.
Monolayer 1$T$ VSe$_2$ shows a CDW, similar to bulk, the periodicity is different.
The bulk CDW has been extensively characterized, and shows a 4$\times$4$\times$3 periodicity in terms of the bulk 1$T$
lattice parameters consistent with the Fermi surface nesting 
\cite{woolley,Strocov_2012,Si_2020}.
The monolayer, on the other hand, has a different normally $\sqrt{7}\times\sqrt{3}$ CDW,
which can be understood in terms
of the Fermi surface nesting \cite{Bonilla_2018,Chen_2018,Feng_2018,Si_2020}.
However, while CDWs are well known, structural distortions in TMDs can have other causes as well,
particularly local bonding instabilities, as in IrTe$_2$
\cite{cao-irte2,fang-irte2}.
Thus the distortions can involve a subtle interplay of Fermi surface effects, as in CDWs, and local
bonding and other chemical effects, potentially leading to complex structures that are not anticipated
based on Fermi surface nesting alone.

An important signature of the CDW in bulk 1$T$-VSe$_2$ is a
reduction in the magnetic susceptibility as temperature is lowered through 
the transition, reflecting partial gapping of the Fermi surface.
Similarly, for the monolayer, standard density functional theory (DFT) calculations
show that the CDW works against magnetism
\cite{Fumega_2019,Coelho_2019}.
However, there are recent reports of CDW states in monolayer
1$T$-VSe$_2$ different from the $\sqrt{7}\times\sqrt{3}$ order mentioned above
\cite{Zong_2021}.
Here we investigate the structure of monolayer 1$T$-VSe$_2$ and find a distortion that leads to
a different metastable polymorph.

\begin{figure}[htbp]
 \centering
 \includegraphics[width=\columnwidth]{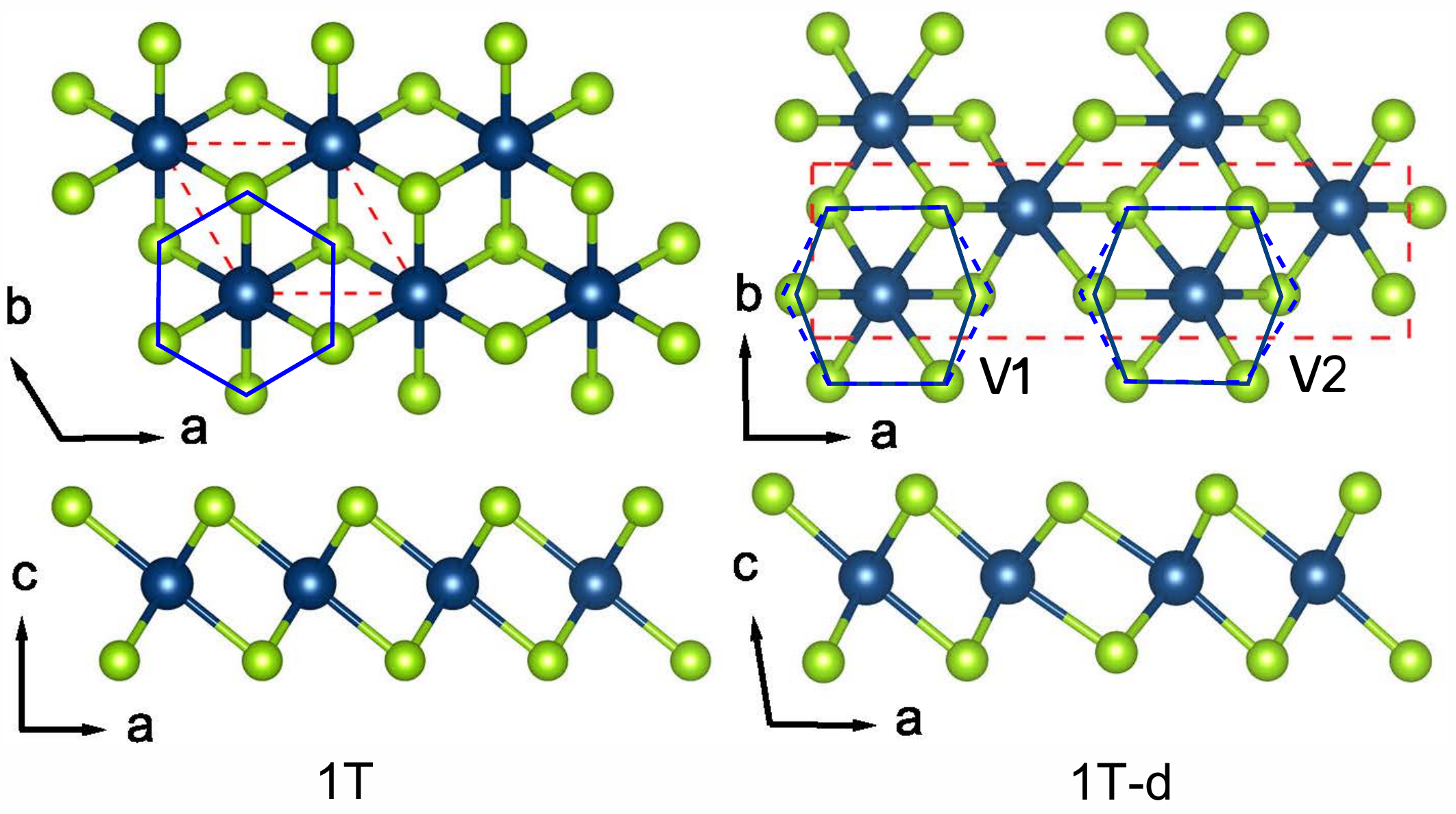}
\caption{\label{fig:fig1}Monolayer structure for the 1$T$ (left) and the new phase 1$T$-d (right).
Note the two
distinct V sites and the asymmetric V environments, especially for the V in the center row of the
top panel.}
\end{figure}

We used DFT calculations to examine instabilities of the 1$T$ monolayer structure, seeking new polymorphs.
For this it was useful to check for instabilities in different supercells.
The motivation for this is the observation that the phonon dispersions for 1$T$-VSe$_2$ apparently
converge slowly with supercell size.
This can be seen for example in the differences of the monolayer dispersions reported
in studies using different cells
\cite{Esters_2017,Coelho_2019,Si_2020}.
This suggests the possibility that a slowly converging
Fermi surface nesting driven instability competes with other instabilities.
The starting 1$T$-VSe$_2$ monolayer structure is shown in Fig. \ref{fig:fig1}.
We used DFT calculations with the projector augmented-wave method \cite{VASP3} implemented in \textsc{vasp} \cite{VASP1,VASP2,VASP4} code with an energy cutoff of 400 eV
and a dense 16$\times$16$\times$1
sampling of the Brillouin zone to study the energy and dynamic stability of this monolayers.
We cross checked results by comparing energy differences obtained with VASP with results from the
all-electron linearized augmented planewave (LAPW) method \cite{LAPW} as implemented in the WIEN2k code \cite{WIEN2k}.
We used to local density approximation (LDA) for the present calculations \cite{LDA},
since it is reported to give good agreement
with the structure of VSe$_2$
\cite{Coelho_2019},
and because it is a conservative approximation from the point of view that it
overestimates magnetic tendencies \cite{tran}.
We did calculations for 1$T$ supercells from  2$\times$2$\times$1 to  6$\times$6$\times$1 in increments of 1,
keeping the lengths of the two directions the same. We then calculated phonon dispersions
with \textsc{phonopy} code \cite{phonopy}.
The zone center density functional perturbation theory \cite{DFPT1,DFPT2} was used for the supercells to obtain the
input for the \textsc{phonopy} calculations.

We find that calculations of the phonons of ideal 1$T$-VSe$_2$ monolayers do not show phonon instabilities
when done using small supercells at and below 3$\times$3$\times$1.
However, once a supercell size of 4$\times$4$\times$1 is reached we do find imaginary frequencies, although the
phonon dispersions remain significantly different from those for larger supercells.
This implies a structural instability, perhaps different from the CDW structure mentioned above.
We did structure relaxations in this 4$\times$4$\times$1 1$T$ supercell starting with random
displacements and relaxed the lattice parameters and atomic positions, while constraining the volume to
preserve the monolayer.
This leads to a structure different from that of the CDW, as shown in Fig. \ref{fig:fig1}.
Symmetry analysis shows that it has the centrosymmetric $P2_1/m$ (No. 11) space group.
The lattice parameters are
$a$=11.11 \AA, $b$=3.22 \AA, $c$=21.30 \AA,
$\alpha$=90$^{\circ}$,  $\beta$=96.15$^{\circ}$, $\gamma$=90$^{\circ}$,
$Z$=4.
The atomic positions are
V1: 0.353, 0.750, 0.000,
V2: 0.114, 0.250, 0.000,
Se1: 0.216, 0.750, 0.077,
Se2: 0.973, 0.250, 0.078,
Se3: 0.729, 0.750, 0.077,
Se4: 0.473, 0.250, 0.068.
We also investigated the phonon dispersions of the distorted 1$T$-d monolayer. This was
obtained from phonon calculations, as above, but with a 3$\times$3$\times$1 supercell of the distorted structure.
Importantly, the structure is found to be dynamically stable. We also checked a smaller 2$\times$2$\times$1 supercell
and find the same result.

The resulting monolayer structure is similar to the undistorted 1$T$ structure
in the sense that the transition metal coordination remains octahedral,
but there are significant differences. Most notable is the fact that the V-Se bonds
in differ in this distorted phase vary from 2.35 \AA -- 2.54 \AA, while in the
normal 1$T$ monolayer structure they are equal, and with LDA relaxed Se positions are 2.47 \AA.
The distortion where one Se is closer to V and where there are rather distinct V sites is evident in
the right panel of Fig. \ref{fig:fig1}.
This asymmetric coordination suggests a Jahn-Teller or other local mechanism for the distortion.
In fact, what we find is a local bonding instability that leads to off-center V, similar to the
vanadyl bond in some V$^{4+}$ oxide systems along with a disproportionation into two chemically different V sites.
This implies a much richer interplay between local bonding and Fermi surface nesting than previously known
for VSe$_2$.
Furthermore, as mentioned there are two distinct V sites in the distorted structure.
The V1 site, which is the more asymmetrically coordinated site, has a calculated bond valence sum \cite{brown,altermatt}
of $BVS$=4.64, while the V2 site has $BVS$=4.33. This is a substantial different, indicative of a partial 
disproportionation. The V in the ideal 1$T$ structure with LDA relaxed Se positions has a lower $BVS$=4.21
indicating more bonding in the 1$T$-d structure.
Calculated energetics, including possible ferromagnetism, are given in Table \ref{tab:table1}
for the 1$T$-d structure in comparison with the undistorted structure, the reported CDW structure and
the alternative 1$H$ structure.
As seen the energies of the 1$T$-d structure and the experimentally observed CDW structure are
the same to within the precision of the calculations, meaning that the 1$T$-d structure is a 
low energy metastable structure that could realistically be found in experiments and that tuning
parameters such as minor strains or chemisorption likely could switch from one structure to the other.

\begin{table}[tb]
\caption{\label{tab:table1}
Calculated energy, E (meV/f.u.) 
and magnetic moment, M ($\mu_B$/f.u.) of ferromagnetic (FM) configurations where they are found.
Data is for the 1$T$, $\sqrt{7}\times\sqrt{3}$ CDW, 2$H$ and distorted 1$T$-d monolayers comparing
results from \textsc{vasp} and WIEN2k.
The energy zero is for the non-spin-polarized (NM) ideal structure 1$T$ monolayer.}
\begin{ruledtabular}
\begin{tabular}{ccccc}
   &\multicolumn{2}{c}{\textsc{vasp}} &\multicolumn{2}{c}{\textsc{wien2k}} \\
\hline
Phase&NM&FM (M)&NM&FM (M)\\
\hline
1$T$& 0.00 &-6.3 (0.46)&0.00&-7.5 (0.47)\\
2$H$& -63.9&-70.5 (0.73)&-64.8&-75.4 (0.82)\\
$\sqrt{7}\times\sqrt{3}$& -101.3&  & -136.5& \\
1$T$-d & -108.3& & -141.8 & \\   
\end{tabular}
\end{ruledtabular}
\end{table}

\begin{figure}[htbp]
 \centering
 \includegraphics[width=0.95\columnwidth]{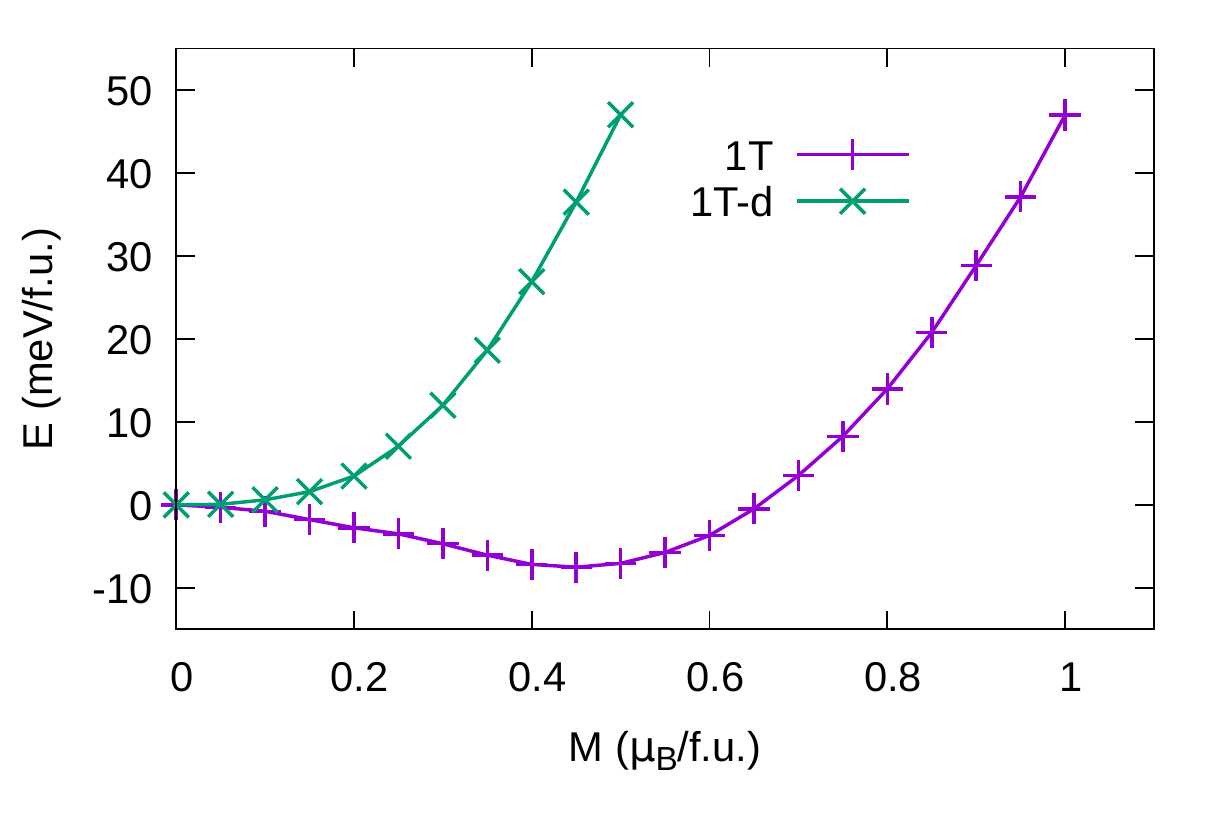}
\caption{\label{fsm}Fixed spin moment total energy for the 1$T$ and 1$T$-d monolayer phases as a functions
of constrained spin magnetization on a per formula unit basis with the LDA.
The energy of the non-spin-polarized case was set to zero.}
\end{figure}

\begin{figure}[htbp]
 \centering
 \includegraphics[width=0.95\columnwidth]{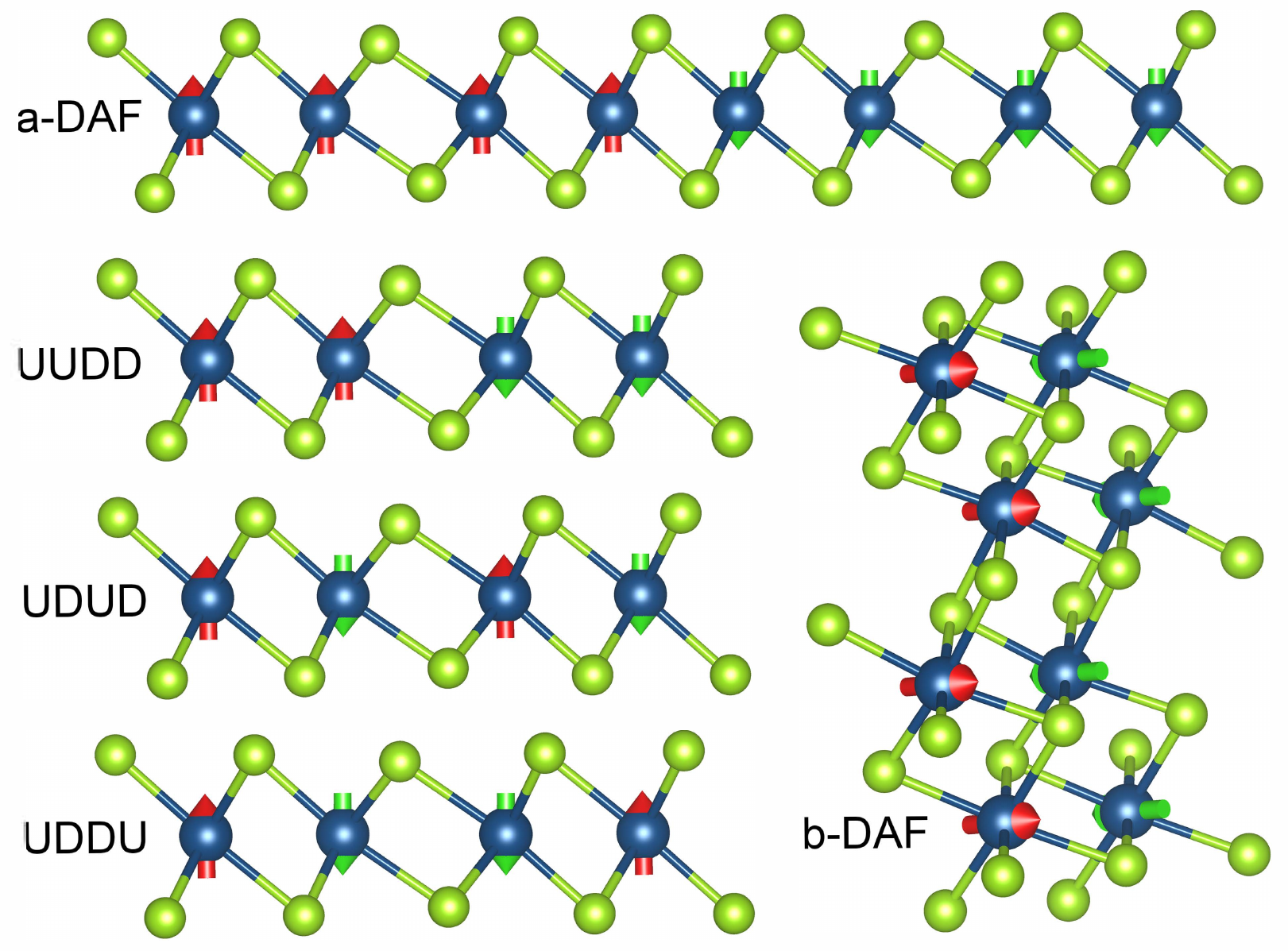}
\caption{\label{fig:fig3}Magnetic configurations investigated for the 1$T$-d phase:
up-up-down-down phase (UUDD), up-down-up-down (UDUD) phase,
up-down-down-up phase (UDDU) phase, antiferromagnetic with doubled cells along the $a$ axis (a-DAF)
and along the $b$ axis (b-DAF).}
\end{figure}

The 1$T$-d phase shows much less tendency towards magnetism than the ideal
undistorted 1$T$ phase.
Calculated energies from constrained density functional fixed spin moment calculations
are given in Fig. \ref{fsm}.
There is no ferromagnetic instability for the 1$T$-d structure, in contrast to the weak ferromagnetic
instability of the ideal undistorted structure.
We investigated various possible magnetic configurations,
including ferromagnetic (FM), as in Table \ref{tab:table1}
and various antiferromagnetic
configurations, as shown in Fig. \ref{fig:fig3}.
For this we used the LAPW method as implemented in the WIEN2k
code \cite{WIEN2k}.
We used well converged basis sets with the convergence criterion $R_{min}k_{max}$ = 9, where $R_{min}$ is the smallest LAPW sphere radius and $k_{max}$ is the planewave sector cutoff.
We included local orbitals for the semicore states.
However, with the LDA none of these configurations yielded a spin-polarized solution, meaning that at the
LDA level the 1$T$-d structure is predicted to be non-magnetic.

\begin{figure}[htbp]
 \centering
 \includegraphics[width=\columnwidth]{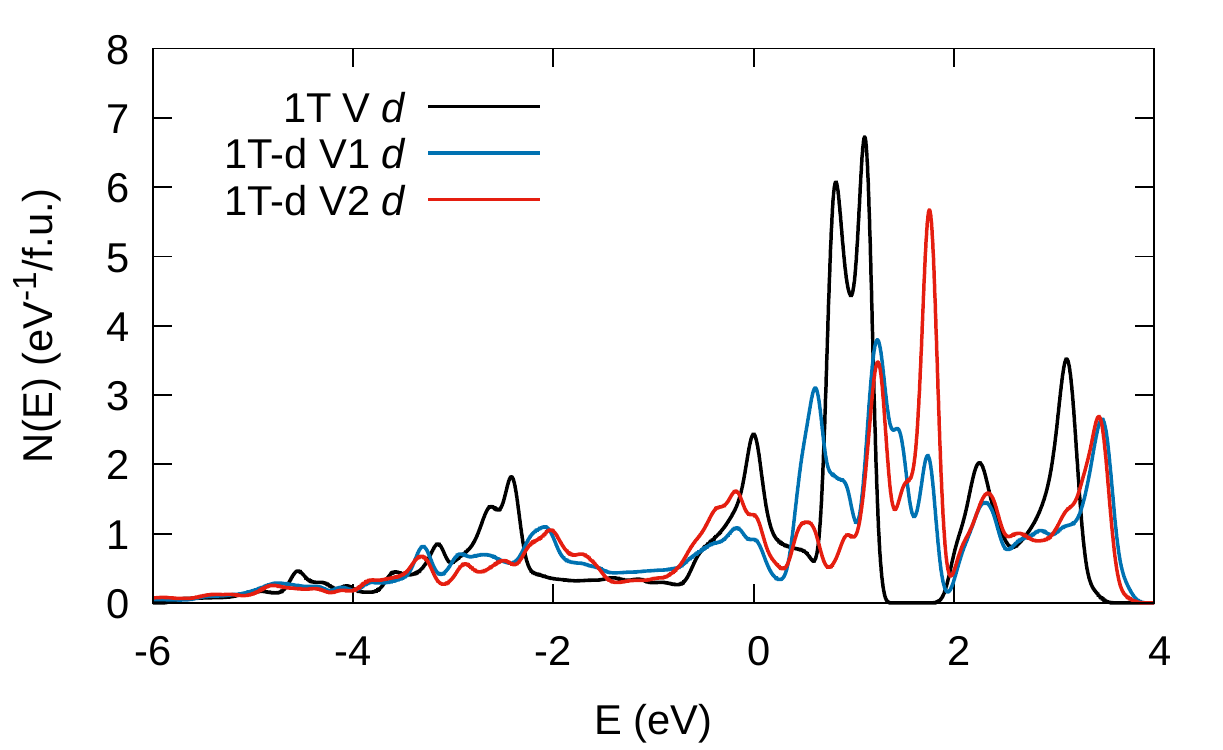}
\caption{\label{fig:fig4}V $d$ projections of the
electronic density of states on a per V basis for the 1$T$ structure and the two different
V sites of the 1$T$-d phase.
The energy zero is at the Fermi level and a Gaussian broadening of 61 meV was applied.}
\end{figure}

CDW instabilities are driven by an energy lowering due to gapping of nested Fermi surfaces.
As discussed in prior work, this is the basic mechanism for the CDW instabilities of bulk and monolayer 1$T$ VSe$_2$
\cite{woolley,Strocov_2012,Si_2020,Bonilla_2018,Chen_2018,Feng_2018,Si_2020}.
The thus reduced density of states (DOS) at the Fermi level, $N(E_F)$ reduces the tendency towards itinerant
Stoner ferromagnetism.
The situation for the 1$T$-d structure is more complex as illustrated by comparison of the projected
V $d$ DOS shown in Fig. \ref{fig:fig4}.
The 1$T$-d structure does have a reduced value of
$N(E_F)$=2.09 eV$^{-1}$ per formula unit,  as compared to $N(E_F)$=5.35 eV$^{-1}$
for the ideal 1$T$ structure.
This explains the absence of a ferromagnetic instability via the Stoner mechanism as seen in Fig. \ref{fsm}.
However, there is strong reconstruction of the DOS throughout the V $d$ bands,
as in IrTe$_2$.
\cite{cao-irte2,fang-irte2}.
For example, there is a shift to higher energy of the top of the $d$ bands at $\sim$3.5 eV.
Such an upward shift in antibonding states indicates stronger bonding interactions, consistent with the bond
valence sums.
In addition, it is important to note that the DOS in the main $d$ bands is significantly different
for the two V sites in the 1$T$-d structure, particularly in the range from the $E_F$ to 2 eV.
In any case, changes extending over the $d$ bands are characteristic of a distortion that is associated
with local bonding effects, as opposed to just a Fermi surface instability.

In summary,
we identify a 4$\times$1 metastable distorted structure of monolayer 1$T$ VSe$_2$.
This structure is non-magnetic and distinct from the known CDW phase.
It has an energy very close to that of the previously identified CDW structure, which implies that it
should be experimentally observable. Importantly, the physics of this distortion is distinct from
that of the CDW in that it is driven at least in part by local bonding instabilities. These local bonding
instabilities
result in a disproportionation into two rather distinct V sites, as well as V off-centering inside
the octahedral Se cages.
This shows that the physics of 1$T$ VSe$_2$ is much richer than previously known, in particular
showing an interplay of Fermi surface driven instabilities, local bonding effects and properties.

\begin{acknowledgments}
This work was supported by the Department of Energy, Basic Energy Sciences, Award DE-SC0019114.
\end{acknowledgments}

\bibliography{VSe2}

\end{document}